# Envisioning Tool Support for Designing Privacy-Aware Internet of Thing Applications

Charith Perera, Mahmoud Barhamgi, Massimo Vecchio

## ABSTRACT

The design and development process for Internet of Things (IoT) applications is more complicated than for desktop, mobile, or web applications. IoT applications require both software and hardware to work together across multiple different types of nodes (e.g., microcontrollers, system-on-chips, mobile phones, miniaturised single-board computers, and cloud platforms) with different capabilities under different conditions. IoT applications typically collect and analyse personal data that can be used to derive sensitive information about individuals. Without proper privacy protections in place, IoT applications could lead to serious privacy violations. Thus far, privacy concerns have not been explicitly considered in software engineering processes when designing and developing IoT applications, partly due to a lack of tools, technologies, and guidance. This paper presents a research vision that argues the importance of developing a privacy-aware IoT application design tool to address the challenges mentioned above. This tool should not only transform IoT application designs into privacy-aware application designs but also validate and verify them. First, we outline how this proposed tool should work in practice and its core functionalities. Then, we identify research challenges and potential directions for developing the proposed tool. We anticipate that this proposed tool will save many engineering hours which engineers would otherwise need to spend on developing privacy expertise and applying it. We also highlight the usefulness of this tool towards privacy education and privacy compliance.

## PRIVACY CHALLENGE AT DESIGN TIME

The engineering complexities in Internet of Things (IoT) have forced engineers to focus most of their efforts on addressing challenges such as interoperability, reliability, and modifiability, resulting in privacy concerns being largely overlooked [1] [2]. IoT applications typically collect and analyse personal data that can be used to derive sensitive information about individuals. Without proper privacy protections in place, IoT applications could lead to serious privacy violations. Over the last few years, we have seen a number of privacy violations (e.g., Baby monitor [3], Google smart speaker eavesdropping[1]). Traditionally, privacy challenges are addressed in an isolated manner by different research communities (e.g., networking, database, software engineering, and human-computer interaction) [1]. More importantly, such independently developed solutions are complicated to adopt and require significant expert knowledge, time, and resources.

In contrast, we propose an end-to-end unified technique that does not require expert knowledge in order for it to be adopted, therefore reducing the cost associated with designing privacy-aware IoT applications. Our vision is to develop a tool that the software engineering community has not seen before that would not only bring privacy-by-design (PbD) into mainstream engineering but also ensures that IoT applications are compliant with leading laws and regulations (e.g., General Data Protection Regulation (GDPR) [4]) working towards creating a safer and privacy-aware IoT ecosystem. Usability, consistency, trustworthiness, scalability, accuracy, accessibility, and extendibility are major unique characteristics of this tool.

---

[1] http://money.cnn.com/2017/10/11/technology/google-home-mini-security-flaw/index.html

Contribution: The primary goal of this paper is to highlight the importance of developing tools to augment the capabilities of software engineers towards designing privacy-aware applications. First, we present an example scenario where we argue the need for a tool. Then, we discuss how such a tool should behave and the functionalities it should provide to the developers. Finally, we provide some insights on how to develop such a tool by linking to existing literature.

**Walking through an example**

Let us consider a simplified use case scenario to highlight the challenges in designing privacy-aware IoT applications. A doctor needs an IoT application which can be used to monitor patients' rehabilitation process. This use case is inspired by a real-world application called *'MyPhysioapp'* (myphysioapp.com) [5]. A doctor has compiled his functional requirements as follows. The doctor has difficulties in seeing his patients frequently due to different reasons (e.g., travelling distance, work schedules, etc.). Further, frequent in-person consultations are not necessary for most circumstances. Each in-person visit incurs costs for both the doctor (government) and the patient. Once the initial consultation is performed, the doctor only needs to track the patient's progress and does not need to meet the patient unless something exceptional has happened. The doctor is only interested in tracking the patient's progress. After evaluating the progress every two weeks, the doctor may ask his speciality nurse to change the exercise plan as necessary. Two software engineers have come up with two different designs as follows to fulfil the above functional requirements. The designs are visually illustrated in Figure 1.

- **Design 1:** In this design, wearable sensors are used to capture raw data (e.g., accelerometers, gyroscopes) that can be used to identify users' (patients) activities. Data is then sent to the cloud for activity recognition using a mobile phone as an intermediary device. Next, cloud services are used to process the raw data and identify the user's activity patterns. User activity patterns are then compared with the doctor's recommended rehabilitation plan to produce a progress report. The doctor can review the progress and make recommendations to the nurse regarding any alterations.

- **Design 2:** In this design, wearable sensors are not only used to capture raw data but also to identify activities (using the micro-controllers attached to the wearable). Timestamped activities are then sent to the mobile phone. The nurse, based on the doctors' recommendations, creates the exercise plan and sends it to the patient's mobile phone. The mobile phone then compares the timestamped activity data and the exercise plan in order to determine how well the patient is performing the exercises. The mobile phone sends a weekly progress report to the doctor. Based on the report, the doctor gives advice to the nurse, and the nurse alters the exercise plan accordingly.

It is important to note that both designs satisfy the doctor's functional requirements. However, design 2 is certainly 'better' than design 1 in terms of privacy awareness. Based on this use case scenario, we extract two research questions as follows:

How can we define and operationalise 'better' IoT application designs (in terms of privacy)

How can we automatically convert a weaker design (e.g. design 1) into a better design (e.g. design 2)

We consider privacy as a trade-off function. Applying a certain privacy-preserving measure into a certain IoT application may impact the implementations in terms of costs, complexity, usability, fault tolerance, responsiveness, etc. Therefore, our aim is not to prescribe a certain design over others. Instead, we want the developer to be informed about privacy-by-design choices before they make their final design decisions. In this regard, we propose a usable privacy-aware IoT application design tool which will inform the privacy-aware design choices to the developers. Previous investigations have shown that applying privacy principles into IoT applications is time-consuming and difficult [6]. The term 'Usable Privacy' means we emphasise on both privacy and usability equally. More specifically, we highlight the importance of putting the user (in this case, the developer) at the centre and consider their requirements, constraints, priorities, and needs when designing a tool.

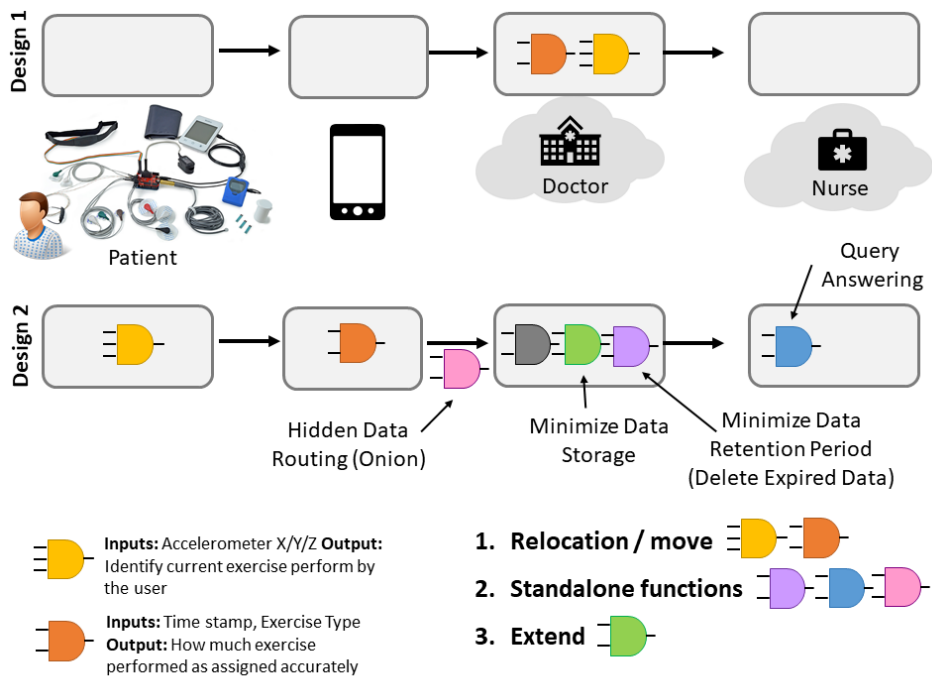

*Figure 1: Motivational Scenario: Different IoT application designs can be developed to fulfil the same functional requirements with different privacy risks associated with them.*

**Target Roles and Audience**

We believe such a tool can be beneficial to different types of stakeholders as follows:

- **Design Tool for Software Engineers (designers/architects):** Primary stakeholders of this tool would be software engineers. We expect them to use this tool to sketch their potential IoT application designs and get validated before moving to the implementation phase. This tool will provide different types of suggestions which engineers can use to improve their IoT application designs in terms of privacy. More importantly, we do not expect this tool to act as a black box that just spits out application designs. Instead, each recommendation will be justified by the tool so engineers can understand why the tool is making a certain recommendation.

- **Compliance Tool:** This tool will also be useful to demonstrate certain compliance needs (e.g., GDPR [4]). It will have the capability to automatically generate a compliance report for each IoT application design briefly explaining the design decisions and risks associated with them so that the compliance officers can determine whether to approve or not. We envision that, in the future, such a compliance tool could be useful (or maybe required) when submitting an IoT application to IoT app stores.

- **Education and Awareness Tool:** We also expect this tool to be used for enhancing privacy awareness among students from school level to university level. Over the last few years, many programming environments, and languages have been developed to help young children to learn how to code (tynker.com, scratch.mit.edu). Similarly, we believe that everyone should learn about privacy at a young age, so over time, they will become responsible software engineers who care about privacy. A Privacy Mindset [6] is really important to be developed by software engineers.

## TOOL-ASSISTED PRIVACY-AWARE IOT APPLICATIONS DESIGN

The tool we propose is something that the engineering community has not seen before. It is inspired by many existing tools used by the engineering community (e.g., UML design tools). First, let us illustrate how the proposed tool (and underlying technology) is expected to work in practice using Figure 2.

**(Step 1)** Software engineers will draw their application designs using a pre-defined set of notations. Key components will be nodes (device profiles) and data flows. To ease the process, common device profiles will be provided. This process will look like a UML diagram design process. **(Step 2)** Engineers will then specify the service which they plan to run. **(Step 3)** They can either assign each service to a node or just leave them unassigned for the algorithms to do that in a later step. **(Step 4/5)** Optionally, engineers can provide additional information related to data management (e.g., 90 days of data retention) and context (e.g., healthcare domain). Additional information will help the algorithms to better design IoT applications. The rest of the steps are invisible to engineers and triggered by a single click. **(Step 6)** Algorithms automatically assign each service into nodes appropriately by considering device capabilities, runtime requirements of the services, and other relevant context information. **(Step 7)** Algorithms incorporate privacy protection features into the design. This step may also reassign the services into different nodes, if necessary. This is one of the key features of this tool. **(Step 8)** Algorithms examine the privacy awareness at both node and composition levels. Then, all the results will be combined to produce the overall privacy index and present it to the engineers. Engineers may consider changing their initial designs to improve the privacy index. **(Step 9)** The terms and conditions unique for each IoT application design are automatically generated.

So far, we designed the IoT application assuming the target environment is static. This works sufficiently for use cases like above. However, some types of applications would require adaptation at run time to better serve the users. In reality, IoT environments are highly dynamic in nature. Therefore, IoT applications should be able to adapt at run time. Towards this direction, we believe this tool should provide simulation capabilities so the engineers can evaluate how their application might adapt at runtime under different circumstances. Currently, several tools are being developed within the context of security, such as OWASP Threat Dragon (threatdragon.org), Microsoft Threat Modeling Tool, IriusRisk (iriusrisk.com), and CAIRIS (cairis.org). However, none of them explicitly focus on privacy.

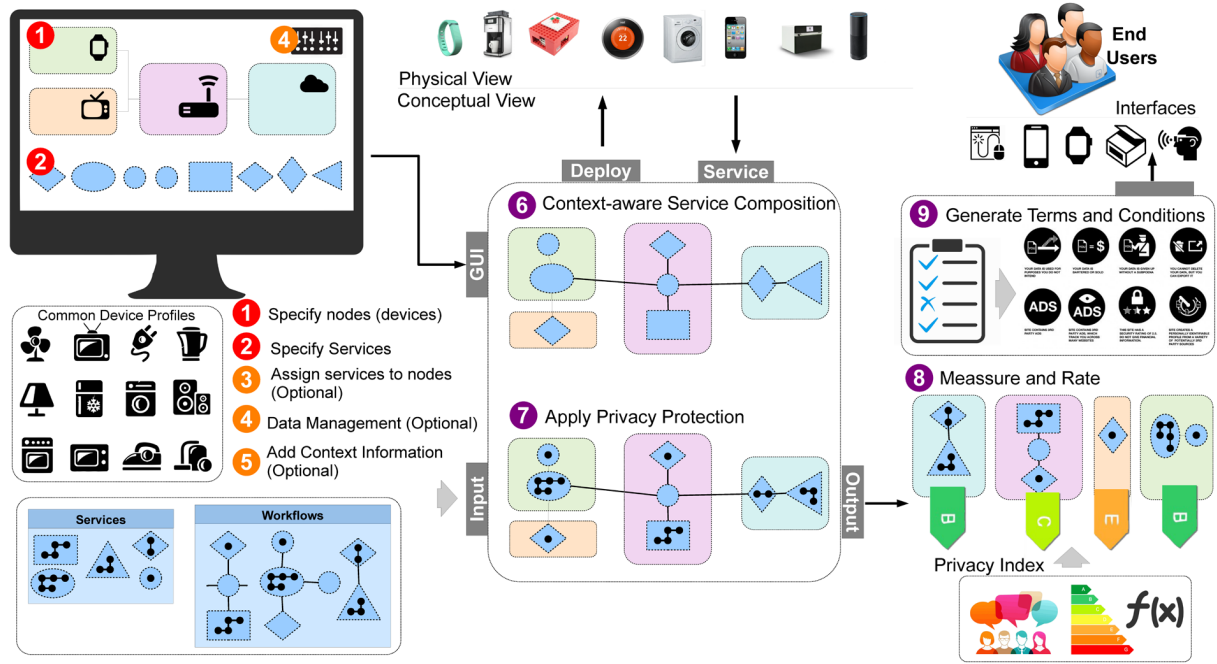

*Figure 2: The Workflow of the Envisioned Tool.*

## Intelligence and Automation

The purpose of the tool is to augment the capabilities of software developers and help them to better design privacy-aware IoT applications. There are many different types of AI techniques which we could be used to develop the tool.

- **Knowledge-based AI:** These techniques can be used to capture the IoT data flow diagrams produced by developers in such a way that they can be processed computationally. For example, class hierarchies and rules (e.g., SWRL) provided within the semantic knowledge engineering domain can be used to conduct reasoning over the graphs. These techniques can be contributed to implementing step 1 to step 9.

- **Case-based reasoning (CBR):** These techniques solve new problems based on the solutions of similar past problems. The steps of CBR are: retrieve, reuse, revise, and retain. These techniques can be used to expedite the step 6 and step 7. For example, without starting from the scratch, algorithms can pick a similar solution to the problem at hand (i.e., IoT design drawn by the developer) from is a library and revise it appropriately.

- **Automated planning and scheduling:** These techniques can be used to implement step 6 and step 7 as well. We can convert the problem of integrating privacy protection measures (into an IoT application design) to an AI constraint solver problem. Then, we can use existing AI planning algorithms and frameworks such as Optaplanner (optaplanner.org) to generate new plans (i.e., new designs).

- **Explainable AI:** For example, step 9 can benefit from explainable AI techniques. Developers are not only interested in the end outcome of the tool (i.e., designs that shows how to apply privacy-protecting measures), but also the reasoning and justifications of the recommendations provided by the tool. It is important to tell the developer why it is better to apply 'minimisation' at the first node than applying it at a later node. Explainability is really important to convince engineers with a significant amount of experience where they do not like to follow a tool without proper reasoning or justification.

## STATE OF THE ART

Our vision lies at the intersection of *IoT*, *software engineering*, and *human-computer interaction,* where the primary focus would be on *usable privacy*.

**Software Engineering:** There are a number of existing frameworks that have been proposed to help elicit privacy requirements and to design privacy capabilities in systems. The original privacy-by-design framework was proposed by Ann Cavoukian [7]. This framework identifies seven foundational principles that should be followed when developing privacy-sensitive applications. Building on the ideas of engineering privacy by architecture versus privacy-by-policy presented by Spiekerman and Cranor [8], Hoepman [9] proposed an approach that identifies eight specific privacy design strategies: minimise, hide, separate, aggregate, inform, control, enforce, and demonstrate. In our previous work [10], we dissected Hoepman's [9] high-level strategies into a more prescriptive granular set of thirty guidelines. The STRIDE framework [11] was developed by Microsoft to help software engineers consider security threats. It is an example of a framework that has been successfully used to build secure software systems by the industry. In a similar vein, LINDDUN [12] is a privacy threat analysis framework that uses data flow diagrams (DFD) to identify privacy threats. LINDDUN focuses on eliminating a set of pre-identified privacy threats using a systematic review of data flow diagrams. The LINDDUN approach requires engineers to draw DFD diagrams and manually evaluate them to identify privacy issues using the instructions provided. This type of approach is useful but requires a significant amount of time, effort and some level of privacy knowledge and expertise. Due to human involvement, it is also prone to errors and lacks consistency, and is, therefore, less trustworthy. All the above-mentioned approaches are not focused on design-time aspects where IoT architectures are highly fluid (change over time) and, therefore, need to support run-time adaptation. Privacy (and security) patterns [13] have also been developed to facilitate privacy-aware application designs for web and mobile domains. Some notable work is done by the EU FP7 funded PRIPARE (privacypatterns.eu) where they have developed privacy patterns and categorised them using Hoepman's [9] eight strategies. Though such patterns are useful, usage is limited due to manual work that needs be done (also requires some expert knowledge) by engineers in order to apply them in real-world application designs.

**Human-Computer Interaction (HCI):** HCI techniques are used to increase the privacy awareness of both engineers and end-users. Luger et al. [14] aim to make emerging European data protection regulations more accessible to the public by using a series of privacy ideation cards. They have extracted 40 design principles by examining EU General Data Protection Regulations. These high-level principles are similar to the guidelines and strategies mentioned before but in a visual format. Based on previous efforts on nutrition, warning, energy, and standardised banking privacy notifications, Privacy Label [15] has been proposed as a method to communicate and understand privacy policies using a visual medium. Terms of service are often too long to read and are ignored by end-users. Terms of Service; Didn't Read (ToS; DR) (tosdr.org) aims to address this issue by introducing a rating and classification scheme. It looks for pre-defined factors in a ToS document to generate a rating. Pribot.org  also addresses the same challenge by developing a smart assistant chatbot that can answer questions regarding privacy policies of different websites. PrivacyGrade.org [16] proposes a method to grade smartphone apps from a privacy perspective by evaluating permission requirements. Privacy Bird (privacybird.org) is a tool that allows end-users to find out what web sites will do with their data by reading privacy policies written in the Platform for Privacy Preferences (P3P) standards [2]. Mozilla has proposed privacy icons [3] which aim to replace (or complement) text-based privacy policies with a visual and consistent set of icons. Naeini et al. [17] have explored privacy expectations and preferences in the IoT domain. Their results show that end-users' privacy expectations vary based on contextual factors. Further, end-users prefer to be informed about data management practices. End-users become comfortable in sharing data when honestly communicated. All the approaches mentioned above consider privacy as an afterthought and look at the problem from a third-party point of view. The similarity is that they all propose mechanisms to look at privacy as an afterthought (i.e. looking at a finished product or a service). Further, none of these approaches focus on IoT applications which are complex in nature and have challenging characteristics such as distributed architecture and unpredictability at runtime.

---

[2] http://www.w3.org/2006/07/privacy-ws/papers/24-preibusch-negotiation-p3p
[3] https://wiki.mozilla.org/Privacy_Icons

## BENEFITS AND IMPACT OF THE TOOL

- **Assisting good actors:** It is important to note that the tool we envision is not aiming to stop bad actors with malicious intent. It is designed to assist good actors in designing better privacy-aware applications. More importantly, we aim to assist software engineers (SMEs), freelance developers, IoT hacker/makers who cannot afford to hire privacy engineering expertise or learn privacy engineering capabilities due to practical cost and time constraints. Further, we do not aim to develop a perfect tool (in the short term) that produces perfect results and eliminates human intervention. We intend to highlight and demonstrate the importance of providing useful and usable tools for good actors to do their job better.

- **On the job privacy education:** We believe that such a tool will help novice (in terms of privacy) software engineers to learn privacy challenges and alternative design by design strategies on-the-go without investing significant time and cost on learning them separately.

- **Facilitating constructive and critical discussion through transparency:** Such a tool will allow software engineers (and other stakeholders) to think and act more transparently. It will provide justification base recommendations to improve given IoT application designs. However, we do not expect software engineers to follow all our recommendations. The job of such a tool is to categorise privacy issues based on critical nature and show alternative corrective mechanisms. It is up to the software engineers to decide which privacy issue to address and which to ignore. Our success is not dependent on whether engineers follow our recommendations 100% of the time. Instead, we measure our success based on the tool's ability to create a transparent process where stakeholders can openly and transparently discuss their designs with a critical eye. We believe such a tool will create a starting point (or a common benchmark) so the stakeholders can initiate useful conversations using that common ground.

- **Economic Impact:** Regulatory entities increasingly introduce new laws and regulations that businesses need to follow. Such adherence requires privacy expertise which often needs to be acquired through consultancy services. Such costs are significant for SMEs due to their limited budget. Our technology will help SMEs to reduce (or eliminate in the long term) privacy consultancy. Privacy-by-design will also help SMEs to reduce costs in the long term (e.g., re-engineering cost, privacy violations and legal costs, and loss of customers). It is important to note that GDPR fines can go up to 20 million Euros or 4 per cent of annual global turnover, whichever of both is highest. Especially, such stiff financial penalties could significantly impact SMEs.

- **Social Impact:** Such a tool can impact society in two different ways. First, it will create a benchmark (i.e., Privacy assessment report) to facilitate privacy by design decisions among engineers. Secondly, the same benchmark (in the Terms and Conditions) will help end-users (i.e., consumers of the IoT applications) to understand and choose IoT products based on transparent and trustworthy T&C.

## RESEARCH DIRECTIONS

## Design Notations and User Interactions

We envision this tool to follow the visual programming paradigm [18]. The proposed tool is expected to be used by engineers to design IoT applications by manipulating program elements graphically. We expect such a design process would be natural for engineers as they are typically familiar with design approaches such as Unified Modelling Language (UML) and Data Flow Diagrams (DFD). Such familiarity will help engineers to familiarise themselves with the tool quickly. Ideally, the visual programming language will be inspired by the data flow diagrams notations. A data flow diagram is a graphical representation of the 'flow' of data through a system (in this case, an IoT system), modelling its process aspects. However, it is important to note that DFDs are flexible enough to be represented in different levels of complexities. Therefore, it would be a fine balance between maintaining simplicity while allowing engineers to design their systems in detail. In addition to the DFD based information, the tool should have ways to gather other related contextual information. The tool should show errors/warnings when vital pieces of information are not provided by the engineers (e.g., data retention period). Knowledge-bases can be used to provide assistance and recommendations for the engineers (typical data retention period based on the domain, data types, applicable laws, and so on) [19].

## Privacy Patterns and Knowledge Modelling

Incorporating privacy-preserving techniques into IoT applications is a complex and time-consuming process [6]. Traditionally, in software engineering, such complexities are handled through introducing design patterns. Design patterns are general repeatable solutions to commonly occurring problems. Design patterns can also speed up the design and development process by providing tested and proven solutions. We believe that this tool should be knowledge-driven. This means that algorithms should not require constant upgrades while the knowledge-bases will grow overtime enabling new features and capabilities. To achieve this, we propose to create a privacy patterns library (by both developing new privacy patterns and organising existing privacy patterns). Ontology-based knowledge models can be developed in order to model the information about each privacy pattern in a common structure. Such a common structure and semantic interoperability allow algorithms to manipulate patterns in a semantically meaningful way. Pattern candidates need to be extensively analysed to find out their characteristics (e.g., usability, complexity, abstractness, relationship to other patterns, and composability) and to categorise them from different perspectives (e.g., functionality and level of granularity). Such an analysis would be vital in the next phase.

## Context-Aware Planning and Adaptation

The design and development of IoT applications require both software and hardware to work together across multiple different types of nodes (e.g., micro-controllers, system-on-chips, mobile phones, miniaturised single-board computers, cloud platforms) with different capabilities under different conditions (e.g., CPU, memory, energy, data communication, knowledge availability, energy limitations, latency tolerance limitations, domain requirements). Therefore, the privacy-preserving techniques that can be applied on a given node vary depending on the context. The question that needs to be answered is 'How do we optimally allocate responsibilities to each node based on the context when designing a privacy-aware IoT application?' First, we need a knowledge base that can be used to reason about different IoT application design choices. Secondly, we need algorithms that can decide which privacy patterns are to be used in different nodes individually and as a whole. Towards this, techniques developed by the web service composition community would be useful. This challenge could be addressed by formulating it as a service composition (with constraints) problem [20].

The IoT ecosystem is highly dynamic in nature. Therefore, IoT applications should be able to adapt to the context changes at run time. However, it is difficult to predict how such adaptation would work at runtime. We believe that the proposed tool should be able to provide engineers with some insights on how their applications would adapt. Let us consider a home care example scenario. Assume three IoT systems are deployed in a home as follows: 1) care receiver wears a smart band that tracks health; 2) a smart bed that can adjust itself, 3) a smart carpet that tracks movements. Assume that none of these IoT systems is originally designed to detect falls and notify caregivers. For example, a novel IoT application may be originally developed to detect falls by using smart band data. The challenge is, how should such an IoT application behave if the smart band fails (e.g., hardware failure)? Can the IoT application adapt and reconfigure itself based on context? (e.g., the IoT application reconfigures itself to re-utilise data from the smart bed and smart carpet to detect falls, instead of the failed wrist band). As the IoT application changes at runtime, privacy-protecting measures may also need to change accordingly to support the adaptation. The tool should be able to simulate scenarios to evaluate the quality of the application design as well as its adaptability.

## Operationalisation, Measuring and Rating

Finally, the challenge is how engineers know, given an IoT application design, whether it is a good design or a bad design (from a privacy perspective). We tend to understand different types of measuring and rating/indexing techniques well (e.g., Body Mass Index, energy ratings, food reference intake and so on). However, no such mechanism is available to measure privacy awareness of IoT applications. We believe such mechanisms (e.g., privacy index) would be increasingly important for both engineers and end-users. Engineers can use such an index to evaluate their own applications to improve their designs iteratively. End-users can use such an index to understand how each IoT applications manages their data. Operationalisation of privacy is a challenging task. There are many factors to take into account when generating a privacy index for a particular application design. Some of the major factors are: 1) privacy patterns used (individually and compositions), 2) order of privacy patterns applied, 3) sensitivity of the data involved, 4) potential risks, and so on. After operationalising, these factors need to combine in a meaningful way. Crowdsourcing techniques [21] may be used to combine expert knowledge and end-user expectations to generate a privacy index. The data flow diagrams drawn by developers can be converted into a semantic network which comprises of nodes, and edges (e.g., semantic web model). We can then develop algorithms to traverse through the graph and determine what kind of privacy-preserving measure are being incorporated into the design and where (e.g., which part of the architecture). A privacy index can be generated based on this information. For example, if 'minimisation' privacy-protecting measure is applied earlier (i.e., within the first node) within the architecture (Design 2 in Figure 1), it will get a higher privacy index. Similarly, Design 1 (in Figure 1) will get a comparatively lower index as it applies 'minimisation' within the third node. Likewise, we can define different types of rules that can be automatically checked to generate a privacy index, which will then be combined into a final privacy index.

Another challenge that goes hand in hand with rating is Terms and Conditions (T&C). It is a well-known fact that end users hardly ever read T&C [15] related to any product or service, besides IoT applications. Typically, T&Cs are written as piles of text and therefore, difficult for a human to understand the most important information within the end-users' short interest and attention span. From engineers' point of view, putting together a T&C document is also a time-consuming task that requires a lot of effort and specialist expertise (e.g., legal professionals) and also less trustworthy (due to human involvement). We propose to capture and model privacy expert knowledge using knowledge-based AI techniques so that the algorithms can eliminate the necessity for privacy experts and related human errors. Such knowledge could be used to automatically generate the (T&C) based on the design of the IoT application.

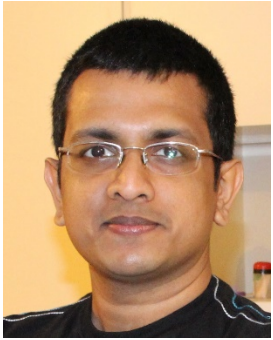

**Charith Perera** (charith.perera@ieee.org) is a Senior Lecturer at Cardiff University, UK. He received his BSc (Hons) in Computer Science from Staffordshire University, UK and MBA in Business Administration from the University of Wales, Cardiff, UK and Ph.D. in Computer Science at The Australian National University, Canberra, Australia. Previously, he worked at the Information Engineering Laboratory, ICT Centre, CSIRO. His research interests are the Internet of Things, Sensing as a Service, Privacy, Middleware Platforms, and Sensing Infrastructure. He is a member of IEEE and ACM. Contact him at www.charithperera.net charith.perera@ieee.org

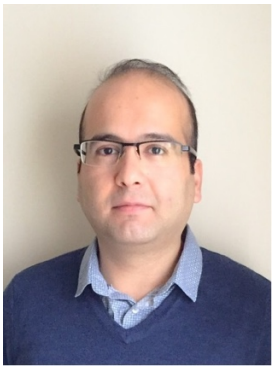

**Mahmoud Barhamgi** (mahmoud.barhamgi@univ-lyon1.fr) is an Associate Professor of computer science at Claude Bernard University Lyon 1, France. His research focuses on security and privacy preservation in service-oriented architecture, web, and cloud environments. Barhamgi received a Ph.D. in information and communication technology from Claude Bernard University Lyon 1. Contact him at mahmoud.barhamgi@univ-lyon1.fr

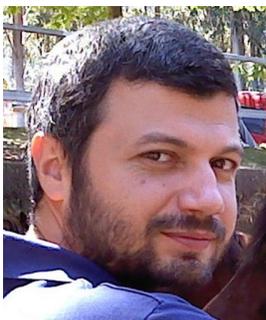

**Massimo Vecchio** (mvecchio@fbk.eu) received his M.Sc. degree in information engineering (magna cum laude) from the University of Pisa, Italy, and his Ph.D. degree in computer science and engineering (with Doctor Europaeus mention) from IMT Institute for Advanced Studies, Lucca, Italy, in 2005 and 2009, respectively. Starting in May 2015, he has been an associate professor at eCampus University, while in September 2017 he also joined FBK CREATE-NET, Trento, Italy, to coordinate the research activities of the OpenIoT Research Unit. He is the project coordinator of AGILE (www.agile-iot.eu), a project co-founded by the Horizon 2020 program of the European Union. His current research interests include computational intelligence and soft computing techniques, the Internet of Things paradigm, and effective engineering design and solutions for constrained and embedded devices. Regarding his most recent editorial activity, he is a member of the Editorial Board of the Applied Soft Computing journal and the Managing Editor of the IEEE IoT newsletters.